\documentclass[aps,prd,groupedaddress]{revtex4-1}
\usepackage{hyperref}
\usepackage{amssymb}
\usepackage{amsmath}
\usepackage{bm}
\usepackage{graphicx}
\usepackage{color}

\def\be{\begin{equation}}
\def\te{\end{equation}}
\def\ee{\end{equation}}
\def\ba{\begin{eqnarray}}
\def\bea{\begin{eqnarray}}
\def\nn{\nonumber\\}
\def\tea{\end{eqnarray}}
\def\ea{\end{eqnarray}}
\def\eea{\end{eqnarray}}

\begin{document}

\title{A Hydrodynamic Approach to the Study of Anisotropic Instabilities in Dissipative Relativistic Plasmas}
\author{Esteban Calzetta}
\email[E-mail me at: ]{calzetta@df.uba.ar}
\affiliation{Departamento de F\'isica, Facultad de Ciencias Exactas y Naturales, Universidad de Buenos Aires and 
IFIBA-CONICET, \\
Cuidad Universitaria, Buenos Aires 1428, Argentina\\
calzetta@df.uba.ar}
\author{Alejandra Kandus}
\email[E-mail me at: ]{kandus@uesc.br}
\affiliation{Departamento de Ci\^encias Exatas e Tecnol\'ogicas, Universidade Estadual de Santa Cruz, \\
Rodov. Jorge Amado km 16, CEP: 45.662-900, Ilh\'eus - BA, Brasil\\
kandus@uesc.br}

\begin{abstract}
We develop a purely hydrodynamic formalism to describe collisional, anisotropic instabilities in 
a relativistic plasma, that are usually described with kinetic theory tools. Our main motivation 
is the fact that coarse-grained models of high particle number systems give more clear and
 comprehensive physical descriptions of those systems than purely kinetic approaches, and can be
 more easily tested experimentally as well as numerically. Also they make 
it easier to follow perturbations from linear to non-linear regimes. In particular, we aim at
 developing a theory that describes both a background 
non-equilibrium fluid configurations and its perturbations, to be able to account for the
 backreaction of the latter on the former. Our system of 
equations includes the usual conservation laws for the energy-momentum tensor and for the electric
 current, and the equations for two new tensors that 
encode the information about dissipation. To make contact with kinetic theory, we write the
 different tensors as the moments of a non-equilibrium 
one-particle distribution function (1pdf) which, for illustrative purposes, we take in the form of
 a Grad-like ansatz. Although this choice limits the 
applicability of the formalism to states not far from equilibrium, it retains the main features of
 the underlying kinetic theory. We assume the validity 
of the Vlasov-Boltzmann equation, with a collision integral given by the Anderson-Witting
 prescription, which is more suitable for highly relativistic 
systems than Marle's (or Bhatnagar, Gross and Krook) form, and derive the conservation laws by
 taking its corresponding moments. We apply our developments to study the emergence 
of instabilities in an anisotropic, but axially symmetric background. For small departures of
 isotropy we find the dispersion relation for normal modes, 
which admit unstable solutions for a wide range of values of the parameter space.
\end{abstract}
\pacs{ 52.27.Ny, 52.35.-g, 47.75.+f, 25.75.-q}

 \maketitle

\section{Introduction}\label{int}
The study of plasma instabilities is of major importance in a wide range of areas as e.g. astrophysics, cosmology, Tokamaks, lasers,  
etc. In the non-relativistic regime, there is a well established hydrodynamic formalism,
 magnetohydrodynamics (MHD),
that consists of the Navier-Stokes equation for the momentum, the continuity equation for the mass
 density and the Maxwell equations for the 
electromagnetic fields, complemented with a corresponding Ohm's law. This theory is known as a
 first order theory, as it is the result of a first 
order expansion in gradients of the distribution function around equilibrium. When turning to
 relativistic domains, it is possible to extend to it the 
tools employed to study ideal fluids, i.e. the Euler equation. But when dissipative processes are
 taken into account, the natural generalization of 
Navier-Stokes equation to relativistic velocities proved to fail, as the solutions are all unstable
 and non-causal \cite{HisLin83}. Among the relativistic 
first order theories, the Eckart \cite{Eck40} and Landau-Lifshitz \cite{LL6} formulations are the
 best known.

Since the seventies several theories were proposed to overcome these drawbacks, among them the 
so-called second order theories
(among several possible strategies \cite{sps}), as e.g. the ones developed by Israel and Stewart
 \cite{SOTs}. Both formalisms, first and second order, 
are based on a gradient expansion of the 1pdf around equilibrium, and in this sense their 
applicability  is limited to small deviations from local equilibrium. There is another set of
 theories, not anchored to a kinetic equation, and that 
are not the result of a perturbative expansion, they are known as Divergence Type Theories (DTT)
 and were developed by Liu and others 
\cite{liu72,LiuMuRu86,GeLin91}. They are exact and thus can describe systems well away from
 equilibrium, but their drawback is that they are not clearly 
linked to microscopic physics.

A paradigmatic case of relativistic plasma is the nuclear matter created in the experiments ongoing
 at the Relativistic Heavy-Ion Collider (RHIC) at 
Brookhaven National Laboratory and at the Large Hadron Collider (LHC) at CERN. There are clear
 experimental signatures that the relativistic matter 
created in the collisions, a quark-gluon plasma, behaves as a strongly coupled system. Consequently
 RHIC's plasmas offer a unique scenario to test 
relativistic hydrodynamics. Indeed pure hydrodynamic models proved to be very successful in
 describing the main features of these plasmas, thus strongly 
improving the understanding of those systems. For a comprehensive review on relativistic
 hydrodynamics and RHIC's plasmas see Ref. 
cite{Sch14,JeHe15,Rom16} and references therein.

RHIC's plasmas show two special features: high degree of anisotropy and quick thermalization. In
 fact, the longitudinal expansion of the fireball causes 
the system to be much colder in the longitudinal direction than in the transverse ones. Such an out
 of equilibrium 
state favors the presence of instabilities and cannot be studied with the usual hydrodynamic models
 based on perturbative schemes around equilibrium 
configurations. Although the development of an anisotropic hydrodynamics \cite{Str14b,TRFS15} is a
 very important step toward the understanding of those 
systems, it is not clear if the resulting hydrodynamics retains enough features of the underlying
 kinetic regime to provide a satisfactory description of 
instabilities. Concerning hydrodinamization, it is believed that the instabilities favored by the
 anisotropic background contribute to such a process. 
Indeed, Mr\'owczy\'nzky showed that they play a substantial r\^ole in the dynamics of the early
 stages of the evolution of quark-gluon plasmas. For more 
details on this issue, see Ref. \cite{MSS16} and references therein. 

There are experimental evidences of the presence of magnetic fields in the RHIC's plasmas 
\cite{Hua16}. Most of the above mentioned anisotropic 
hydrodynamic models do not take into account electromagnetic fields and consequently are not
 suitable to give a realistic explanation of the observations.

The main purpose of our study is to start building a consistent magnetohydrodynamic theory to
 describe strongly coupled, high energy plasmas, without 
having to address to kinetic theory for each different system under study, and that includes both
 the background and its perturbations in a consistent way. 

To facilitate the calculations we consider a massless Abelian plasma. Although a non-Abelian theory
 is needed to correctly describe the plasmas created 
at RHICs, our choice has the advantage of simplifying the mathematics without depriving the model
 of physical relevance 
\cite{MaMr06,PRC12a,PRC09,PRC10b,MSS16}. 

To give our model a kinetic theory support, we write the different tensors as momenta of a
 distribution function and to obtain their
evolution equations we invoke a mean field kinetic model described by the Boltzmann-Vlasov
 equation. As for the collision integral, most of the literature 
uses the BGK (Bhatnagar, Gross and Krook) relaxation time model, as it allows to effectively 
handle distributions other than the Maxwell-Boltzmann. Its 
relativistic generalization was developed by Marle and is
of the form $C(f) = - m(f - f_0)/\tau$ \cite{marle1,marle2}. In the classical limit, Marle's
 formulation gives the same result for the transport 
coefficients as the classical BGK model. However, in the extreme relativistic limit the results for
 the transport coefficients with Marle's formulation 
differ functionally from the ones calculated with the relativistic Grad moment method. Anderson and
 Witting \cite{and-witt1,and-witt2,TI10} proposed an 
improvement of Marle's collision integral of the form $C(f) = - u_{\mu}p^{\mu}(f - f_0)/\tau$ with
 $u_{\mu}$ the four velocity of the gas. 
In the classical limit this expression gives the same classical results as Marle's, since in that
 limit $u_{\mu}p^{\mu}\rightarrow m$, and in the extreme 
relativistic limit it produces the same transport coefficients that are obtained via the relativistic Grad moment method. Consequently, as we are dealing 
with a highly relativistic system we shall adopt the Anderson-Witting prescription instead of the
 BGK collision kernel in Marle's form, generally adopted 
in the literature.

We consider a model where the distribution function is the product of an equilibrium expression
 times a non-equilibrium part. The former is isotropic 
and homogeneous in the momenta, and also depends on a thermal potential that accounts for possible
 excess of particles over antiparticles. We specify it 
by demanding that the ideal energy-momentum tensor $T^{\mu\nu}$ calculated from it, corresponds to
 the  Landau-Lifshitz prescription, whereby in the rest 
frame $T^{0i}=0$ \cite{LL6}. The latter contains all the information about anisotropies and
 dissipation. For the collision integral, we only demand it to 
be linear in the tensors that describe non-equilibrium features, i.e., ohmic and viscous
 dissipation, and possible anisotropies in the momenta 
distribution. The main motivation behind this choice is to avoid mathematical complexity.

Our model is not truly reliable for arbitrarily large anisotropy, as it will be discussed below
 (see also Ref. \cite{MAEC17}). Within its range of 
validity, however, it fully captures nonlinearities coming from the convective derivative terms and
 from direct coupling of the hydrodynamic variables 
to the electromagnetic fields in the equations of motion. These are the only nonlinearities in the
 usual magnetohydrodynamics, where dissipative terms 
are assumed to be linear. For this reason, we believe the hydrodynamic equations to be introduced
 below (eqs. (\ref{b33}) to (\ref{b43b})) are a valid 
generalization of MHD to the relativistic regime. Moreover, we also believe any consistent
 relativistic dynamics of real fluids will converge to this 
formalism within its range of validity.

We build our formalism by writing the different tensors as moments of the distribution function,
 and find their evolution equations by taking the 
corresponding moments of the Vlasov-Boltzmann equation. By projecting those equations along the
 four velocity and onto its orthogonal hypersurface we 
obtain five hydrodynamic equations: for the charge density, for the energy density, for the
 velocity field and for the two tensors that describe 
dissipation. Together with Maxwell equations they form our magnetohydrodynamic model.

As an application of our formalism we study the transverse instabilities that appear in
 fluctuations around an anisotropic background. 
These were first discussed by E. S. Weibel \cite{weibel59} in a non-relativistic setting, and then
 in the relativistic regime in Refs. 
\cite{sschl04,ach1,ach2} among others. In the non-relativistic theory a purely macroscopic
 approach already exists, see e.g. Refs. 
\cite{basu02,brdeu06,bret06}. Our aim is to generalize these macroscopic approaches to
 relativistic theories, accounting for dissipative effects. 
Of course the Weibel instability is not the only possible instability of relativistic plasmas, see
 Ref. \cite{bret09} for a detailed analysis of 
the different kinds of instabilities in Abelian plasmas. Moreover when considering non-Abelian
 plasmas new kinds of instabilities appear, as can be 
seen in e.g. Ref. \cite{mama07}. In order to avoid a heavy mathematical content, we leave for
 forthcoming manuscripts the analysis of the other 
instabilities in Abelian plasmas, as well as the extension of our formalism to the non-Abelian 
case.

The manuscript is organized as follows, in Section \ref{kt} we build the 1pdf. In Section \ref{bh}
 we build the magnetohydrodynamic formalism, 
by deducing the tensors and the equations they must satisfy. In Section \ref{pt} we linearize the
 previously found equations around a background with 
anisotropic pressure, and find the dispersion relation for the normal modes, consistent with the
 limitations of the model. For a wide range of values 
of the parameter space, our model predicts the excitation of instabilities, whose features are in
 agreement with results previously found in the 
literature. To illustrate those the dependence on the different parameters of the model, we plot
 this relation for several values of them. Finally, 
in Section \ref{cl} we summarize our conclusions and comment on future perspectives. We work with
 natural units, i.e., $c=\hbar=k_B=1$ and with 
signature $\left(-,+,+,+\right)$.

\section{Kinetic Theory}\label{kt}
In this section we shortly review some basics of kinetic theory of plasmas and build the 1pdf of
 our model. In the mean field approach the kinetic equation for a plasma with electromagnetic
 fields is the Boltzmann-Vlasov equation, which reads
\begin{equation} 
p^{\mu}\left[\frac{\partial}{\partial x^{\mu}}-eF_{\mu\rho}\frac{\partial}{\partial p_{\rho}}\right]f\left(x^{\mu},p^{\mu}\right)=
I_{col}\left(x^{\mu},p^{\mu}\right)\label{a1}
\end{equation}
where $f\left(x^{\mu},p^{\mu}\right)$ is the distribution function, 
$I_{col}\left(x^{\mu},p^{\mu}\right)$ the collision integral (to be defined below). 
Integration over momentum is done with the invariant volume element
\begin{equation}
Dp=\frac{2d^4p}{\left(2\pi\right)^3}\delta\left(p^2\right)=\frac{d^4p}{\left(2\pi\right)^3p}
\left[\delta\left(p^0-p\right)
+\delta\left(p^0+p\right)\right]\label{a2}
\end{equation}
As stated in the Introduction, we shall deal with the massless case, whereby $p^2 = 0$, with $p^0$
 having either sign: positive for positively charged particles, and negative for negatively 
charged antiparticles.

The current and the matter energy momentum tensor (EMT for short) are defined as usual, namely
\begin{equation}
J^{\mu}=e\int Dp\;p^{\mu}f\label{a3}
\end{equation}
and
\begin{equation}
T^{\mu\nu}=\int Dp\;p^{\mu}p^{\nu}f\label{a4}
\end{equation}

$F_{\mu\nu}$ in eq. (\ref{a1}) is the Maxwell tensor $F_{\mu\nu}=\partial_{\mu}A_{\nu}-
\partial_{\nu}A_{\mu}$, with $A^{\mu}$ the electromagnetic 
four potential. Inclusion of the Maxwell field as an independent degree of freedom is of course 
the main goal of this analysis. The Maxwell field 
obeys Maxwell's equations sourced by the current $J^{\mu}$ defined in eq. (\ref{a3})

\bea
F_{\mu\nu,\rho}+F_{\nu\rho,\mu}+F_{\rho\mu,\nu}&=&0\nn
F^{\mu\nu}_{,\nu}&=&4\pi J^{\mu}
\label{meq}
\tea
Antisymmetry of $F_{\mu\nu}$ demands charge conservation

\begin{equation}
 J^{\mu}_{,\mu}=0\label{a6}
\end{equation}
Associated to the Maxwell field there is an electromagnetic energy momentum tensor 
$T_{EM}^{\mu\nu}$ [\cite{MTW}] 

\be 
T_{EM}^{\mu\nu}=\frac1{4\pi}\left\{F^{\mu\rho}F^{\nu}_{\;\rho}-
\frac14\eta^{\mu\nu}F^{\rho\sigma}F_{\rho\sigma}\right\}
\te
$\eta^{\mu\nu}$ is Minkowsky metric. The full energy momentum tensor 
$T_F^{\mu\nu}=T^{\mu\nu}+T_{EM}^{\mu\nu}$ is conserved: $T^{\mu\nu}_{F,\nu}=0$. 
We may also use Maxwell's equations to compute $T^{\mu\nu}_{EM,\nu}=-F^{\mu\rho}J_{\rho}$ 
and thus rewrite the conservation law as

\begin{equation}
T^{\mu\nu}_{,\mu}=F^{\nu\mu}J_{\mu} \label{a7}
\end{equation}
The conservation laws (\ref{a6}) and (\ref{a7}) may be obtained from the zeroth and first 
momenta of the Boltzmann equation, provided that
\begin{equation}
\int Dp\;I_{col}=\int Dp\;p^{\nu}I_{col} =0\label{a5}
\end{equation}

We consider a classical (i.e., not quantum) system. Then in an equilibrium state the 
distribution function takes the form 
\begin{equation}
f_{eq}=e^{\mathrm{sign}\left[p^0\right]\left(\alpha +\beta_{\mu}p^{\mu}\right)}  \label{a8}
\end{equation}
and the collision integral vanishes. In the previous expression $\beta_{\mu}=\beta u_{\mu}$, 
$\beta =1/T$, where $T$ is the temperature. Following Israel [\cite{SOTs}], we call $\alpha$ 
the thermal potential; $\mu =T\alpha$ is the chemical potential 
that accounts for the excess of particles over antiparticles. We choose to identify the 
velocity $u^{\mu}$ and energy density $\rho$ as the timelike 
eigenvector of $T^{\mu\nu}$ and its eigenvalue, i.e., $T^{\mu\nu}_{id}u_{\nu}=-\rho u^{\mu}$, 
i.e., we work in the Landau-Lifshitz frame
\cite{LL6}. Also we define the charge density as $\rho_q=-u_{\mu}J^{\mu}$. In this case the 
ideal part of the current and EMT take the form
\begin{equation}
J^{\mu}_{id}=\rho_q u^{\mu} \label{a9}
\end{equation}
and
\begin{equation}
T^{\mu\nu}_{id}=\rho\left[u^{\mu}u^{\nu}+\frac13 h^{\mu\nu}\right]\label{a10}
\end{equation}
where the fluid four velocity is normalized as $u^2=-1$ and $h^{\mu\nu}=
\eta^{\mu\nu}+u^{\mu}u^{\nu}$ is the projector onto hypersurfaces orthogonal 
to $u^{\mu}$. After evaluating the current and the EMT we read
\begin{equation} 
\rho_q=\frac{2eT^3}{\pi^2}\sinh\alpha\label{a11}
\end{equation}
and
\begin{equation}
\rho=\frac{6T^4}{\pi^2}\cosh\alpha\label{a12}
\end{equation}
Eq. (\ref{a11}) shows that when $\alpha =0$, the number of particles equals the number of
 antiparticles and consequently the net charge of the plasma 
is zero. Moreover expr. (\ref{a11}) and (\ref{a12}) show that the temperature $T$ and thermal
 potential $\alpha$ are univocally determined by the charge and 
energy densities. 

Observe that we may also introduce electric and magnetic fields relative to the fluid rest 
frame by writing

\be
F^{\mu\nu}=u^{\mu}E^{\nu}-E^{\mu}u^{\nu}+\epsilon^{\mu\nu\rho\sigma}B_{\rho}u_{\sigma}
\te
Thus in the rest frame $E^a=F^{0a}$ and $B_a=\left(1/2\right)\epsilon_{abc}F^{bc}$.

To describe non-equilibrium states, we choose to parametrize the distribution function in the 
form:
\begin{equation}
f=f_{eq}\left[1+Z\right]\label{a13}
\end{equation}
We demand that $Z$ satisfies the constraints
\begin{equation}
\int Dp\;f_{eq}\;u_{\mu}p^{\mu} Z=\int Dp\;f_{eq}\;u_{\mu}p^{\mu}p^{\nu} Z=0\label{a14}
\end{equation}
which implies that the ideal forms (\ref{a9}) and (\ref{a10}), with (\ref{a11}) and (\ref{a12}),
 are preserved.
Note that in expression (\ref{a13}), $Z$ is not small in front of $1$. We define the entropy 
flux in the usual way, i.e.,
\begin{equation}
S^{\mu}=-\int Dp\;\mathrm{sign}\left[p^0\right]p^{\mu}f\left[\ln f-1\right]\label{a15}
\end{equation}
which satisfies the equation
\begin{equation}
S^{\mu}_{,\mu}=-\int Dp\;\mathrm{sign}\left[p^0\right]\; I_{col}\ln f\label{a16}
\end{equation}
showing that there is no entropy production from an equilibrium state. To enforce positiveness 
of expression (\ref{a16}), we must choose an 
appropriate collision integral. In this manuscript we concentrate in writing down a simplest
 possible dissipative relativistic 
magnetohydrodynamic formalism to describe high energy plasma features (specially its 
instabilities) without having to resource to kinetic theory 
for each specific problem. The straightforward way to do this is to linearize expr. (\ref{a16}) 
to first order in $Z$, i.e., to write
\begin{equation}
S^{\mu}_{,\mu}=-\int Dp\;\mathrm{sign}\left[p^0\right]Z\;I_{col}\label{a16-b}
\end{equation}
This expression suggests to consider a collision integral of the Anderson-Witting form 
\cite{and-witt1,and-witt2,TI10}, namely
\begin{equation}
I_{col}=\frac{u_{\rho}p^{\rho}}{\tau}f_{eq}Z\label{a17}
\end{equation}
with $\tau$ a relaxation time. This form of $I_{col}$ guarantees that the $H$ theorem is 
satisfied, namely:
\begin{equation}
S^{\mu}_{,\mu}=\int Dp\;\frac{\left|u_{\rho}p^{\rho}\right|}{\tau}f_{eq}Z^2\ge 0\label{a18}
\end{equation}
as well as constraints (\ref{a5}).
To account to dissipative processes in the dynamics we split the electric current and the 
EMT as
\begin{equation}
J^{\mu}=\rho_q u^{\mu}+j^{\mu}\label{a19}
\end{equation}
and
\begin{equation}
T^{\mu\nu}=\rho\left[u^{\mu}u^{\nu}+\frac13 h^{\mu\nu}\right]+\Pi^{\mu\nu}\label{a20}
\end{equation}
where 
\begin{equation}
j^{\mu} = e\int Dp~ p^{\mu}f_{eq}Z\label{a19-b}
\end{equation}
and 
\begin{equation}
\Pi^{\mu\nu} = \int Dp~ p^{\mu}p^{\nu}f_{eq}Z\label{a20-b}
\end{equation}
describe dissipative effects. 

At this point it is necessary to provide an explicit form for $Z$, such that the dissipative 
parts of the current and EMT may be computed. 
Since the EMT is traceless, both conserved currents amount to $13$ degrees of freedom, of 
which $\alpha$, $T$ and $u^{\mu}$ account for $5$. 
It is natural to assume that $Z$ depends on $8$ additional parameters, to which we must 
add $5$ more to have enough freedom to enforce the 
constraints (\ref{a14}). We arrive at the right  number if $Z$ depends on a new vector 
field $Z^{\mu}$ and a tensor field $Z^{\mu\nu}$ such that 
$u_{\mu}u_{\nu}Z^{\mu\nu}=0$. We further split them in longitudinal and transverse components 
along $u^{\mu}$: $Z^{\mu}=e\zeta^{\mu}+a u^{\mu}$ 
and $Z^{\mu\nu}=\zeta^{\mu\nu}+b^{\mu}u^{\nu}+u^{\mu}b^{\nu}$. 
The simplest Lorentz invariant form for $Z$ is the linear one

\begin{equation}
Z=\frac{\tau}{2\left|u_{\rho}p^{\rho}\right|}\left[e\zeta_{\rho}p^{\rho}+
\zeta_{\rho\sigma}p^{\rho}p^{\sigma}+au_{\rho}p^{\rho}
+\frac12\left(b_{\rho}u_{\sigma}+u_{\rho}b_{\sigma}\right)p^{\rho}p^{\sigma}\right]
\label{a22}
\end{equation}
hence $S^{\mu}_{;\mu}\geq 0$. Since $p^{\mu}$ is restricted to the null cone, we may impose 
one further condition on $\zeta^{\mu\nu}$: 
we chose it to be traceless. The functional form (\ref{a22}) can also be obtained by 
using a variational method to impose  
constraints that describe the non-equilibrium state of the system, such as the Entropy 
Production Variational Method 
\cite{epvm-rev,christen14,christen,christen2}. It can be proved that in out-of-equilibrium 
linear thermodynamics, stationary states are extrema of 
the entropy production rate. Moreover, at linear order in the entropy production, the results 
are equivalent to those obtained through the Grad 
approach \cite{PRC10a,PRC12a,PRC13a,epv-2}.

Recalling that $\zeta_{\rho}$ and $\zeta_{\rho\sigma}$ are transverse and the latter is 
traceless, constraints (\ref{a14}) read
\begin{eqnarray}
0&=&\cosh\alpha\; a+3T\sinh\alpha\;b_{\rho}u^{\rho}\nonumber\\
0&=&\sinh\alpha\left[-\frac e3\zeta^{\nu}+au^{\nu}\right]+4T\cosh\alpha\;b_{\rho}
\left(u^{\nu}u^{\rho}
+\frac13 h^{\nu\rho}\right)\label{a25}
\end{eqnarray}
whose solutions are
\begin{eqnarray}
a&=&b_{\rho}u^{\rho}=0\label{a26}\\
b^{\nu}&=&\frac e{4T}\tanh\alpha\;\zeta^{\nu}\label{a27}
\end{eqnarray}
Replacing in eq. (\ref{a22}) we finally obtain
\begin{equation}
Z=\frac{\tau}{2\left|u_{\rho}p^{\rho}\right|}\left[e\zeta_{\rho}p^{\rho}+\zeta_{\rho\sigma}
p^{\rho}p^{\sigma}
+\frac e{4T}\tanh\alpha\; \zeta_{\rho}u_{\sigma}p^{\rho}p^{\sigma}\right]\label{a22-b}
\end{equation}
The tensors $\zeta^{\mu}$ and $\zeta^{\mu\nu}$ are the new ones mentioned in the Introduction. 
They account for the different dissipative processes: 
the former represents conduction currents, while the latter is associated to viscous 
stresses.

\section{Building the Hydrodynamics}\label{bh}

The different tensors that describe our hydrodynamical model are written in terms of 
the distribution function in the usual way, namely
\begin{equation}
A^{\mu_1,\ldots\mu_n}_{s}=\int Dp\;\left(\mathrm{sign}\left[p^0\right]\right)^sp^{\mu_1}
\ldots p^{\mu_n}f \label{b1}
\end{equation}
with $s=0$ or $1$. The conservation laws obeyed by these tensors are obtained by taking the corresponding moments of eq. (\ref{a1}), 
their general form then being
\begin{equation}
A^{\mu\mu_1,\ldots\mu_n}_{s,\mu}-e\sum_{i=1}^nF^{\mu_i}_{\;\mu}A^{\mu\mu_1,\ldots
\left(\mu_i\right)\ldots\mu_n}_{s} 
=-I_s^{\mu_1,\ldots\mu_n}\label{b2}
\end{equation}
where the notation $A^{\mu\mu_1,\ldots\left(\mu_i\right)\ldots\mu_n}_{A}$ means that 
$\mu_i$ is excluded, and 
\begin{equation}
I_s^{\mu_1,\ldots\mu_n}=-\int Dp\;\left(\mathrm{sign}\left[p^0\right]\right)^sp^{\mu_1}
\ldots p^{\mu_n}I_{col}\label{b3}
\end{equation}
Each momentum may be written as $A^{\mu_1,\ldots\mu_n}_{s}=A^{\mu_1,\ldots\mu_n}_{s,ideal}
+A^{\mu_1,\ldots\mu_n}_{s,dis}$ with
$A^{\mu_1,\ldots\mu_n}_{s,ideal}$, $A^{\mu_1,\ldots\mu_n}_{s,dis}$ and 
$I_s^{\mu_1,\ldots\mu_n}$ totally symmetric and traceless 
on any two indices.

From expr. (\ref{b1}) we thus obtain the different tensors of our model; in particular, 
the current previously introduced in eq. (\ref{a19}) is $J^{\mu}=eA^{\mu}_0$ and the EMT 
defined in eq. (\ref{a20}) is $T^{\mu\nu}=A^{\mu\nu}_0$. As discussed above, our theory has 
$13$ non trivial degrees of freedom $\alpha$, $T$, $u^{\mu}$, $\zeta^{\mu}$ and 
$\zeta^{\mu\nu}$. The charge and EMT conservation laws provide $5$ equations. 
To obtain the necessary $8$ supplementary equations we will consider two more tensors 
$A^{\mu\nu}_1$ and $A^{\mu\nu\rho}_1$. The equations we seek are $h_{\mu\lambda}
\left[A^{\nu\lambda}_{1,\nu}+I^{\lambda}_1\right]=0$ and $\left(h_{\mu\lambda}h_{\nu\sigma}
-\left(1/3\right)h_{\mu\nu}h_{\lambda\sigma}\right)\left[A^{\rho\lambda\sigma}_{1,\rho}
+I^{\lambda\sigma}_1\right]=0$. The former provides $3$ new equations, and the latter the 
remaining $5$.

Let us now compute the relevant tensors. At first level:
\begin{equation}
A^{\mu}_{0}=q_1T^3 u^{\mu}+\Lambda e\tau T^3\zeta^{\mu}\label{b7}
\end{equation} 
where $q_1=\left(2/\pi^2\right)\sinh\alpha$, $q_2=\left(2/\pi^2\right)\cosh\alpha$ and 
$\Lambda = \left(4q_2^2-3q_1^2\right)/24q_2$. 

The vector $A^{\mu}_{1}$ is the particle number current, which in our model is likewise 
conserved. However, this is actually a drawback of the model, which is too simplistic to 
account for pair creation and annihilation. We therefore pass it over and consider other 
currents whose conservation laws may be expected to be less sensitive to those effects. 
In other words, while charge and EMT conservation hold for any form of the collision 
integral, as long as the constraints eq. (\ref{a5}) are enforced, the conservation laws 
we are writing down for the other tensors depend on the precise form of the collision 
integral. In this sense, we may regard the Anderson - Witting (AW) collision integral 
as a first order approximation in a series expansion in which progressively more complex interactions are taken into account. The particle number current is highly sensitive to 
the higher order terms in this expansion, because in this case the production term computed 
from the AW collision integral vanishes; therefore the first order equation is not reliable. 
For $A^{\mu\nu}_1$ and $A^{\mu\nu\rho}_1$, as we shall see presently, the AW collision 
integral gives nontrivial production terms, and so the dependence on further improvements 
of the collision integral may be expected to be weaker.

At second level:
\begin{equation}
A^{\mu\nu}_{0}=3q_2T^4\left[u^{\mu}u^{\nu}+\frac13 h^{\mu\nu}\right]+\frac{4 q_2 }{5}\tau
 T^5\zeta^{\mu\nu} 
 \label{b9}
\end{equation}
\begin{equation}
A^{\mu\nu}_{1}=3q_1T^4\left[u^{\mu}u^{\nu}+\frac13 h^{\mu\nu}\right]
+\kappa_1 e\tau T^4\left[\zeta^{\mu}u^{\nu}+\zeta^{\nu}u^{\mu}\right]+\eta_1 \tau 
T^5\zeta^{\mu\nu}\label{b10}
\end{equation}
with $\eta_0=4 q_2/5$, $\eta_1=4 q_1/5$ and $ \kappa_1=\left(q_2^2-q_1^2\right)/{2q_2}$. 
Finally, at third level
\begin{eqnarray}
A^{\mu\nu\rho}_{1}&=&12q_2T^5\left[u^{\mu}u^{\nu}u^{\rho}+\frac13\left(h^{\mu\nu}u^{\rho}
+h^{\mu\rho}u^{\nu}+h^{\rho\nu}u^{\mu}\right)\right]\nonumber\\
&-&\frac{q_1}{2} e\tau T^5\left[\zeta^{\mu}u^{\nu}u^{\rho}+\zeta^{\nu}u^{\mu}u^{\rho}
+\zeta^{\rho}u^{\nu}u^{\mu}
+\frac15\left(h^{\mu\nu}\zeta^{\rho}+h^{\mu\rho}\zeta^{\nu}
+h^{\rho\nu}\zeta^{\mu}\right)\right]\nonumber\\
&+& 4\tau q_2 T^6\left(\zeta^{\mu\nu}u^{\rho}+\zeta^{\mu\rho}u^{\nu}
+\zeta^{\rho\nu}u^{\mu}\right)\label{b12}
\end{eqnarray}
There remains to compute the momenta of the collision integral. To do that we observe 
that
$I_s^{\mu_1,\ldots\mu_n}=-\frac1{\tau}u_{\mu}A_{s,dis}^{\mu\mu_1,\ldots\mu_n}$
and therefore
\begin{eqnarray}
I_0&=&I_1=I_0^{\mu}=0\label{b26}\\
I_1^{\nu}&=&e\kappa_1T^4\zeta^{\nu}\label{b28}\\
I_1^{\nu\rho}&=&-\frac{q_1}{2}e T^5\left(\zeta^{\nu}u^{\rho}+\zeta^{\rho}u^{\nu}\right)
+ 4q_2 T^6\zeta^{\nu\rho}\label{b29-b}
\end{eqnarray}

Our hydrodynamic equations for $A_0^{\mu}$, $A_0^{\mu\nu}$, $A_1^{\mu\nu}$ and 
$A_1^{\mu\nu\rho}$ are extracted from the ones obtained 
from (\ref{b2}), by projecting them along $u^{\mu}$ and onto the surfaces defined by 
$h^{\mu\nu}$. We define the new variables 
$q_0=q_1T^3$, $L_0=\Lambda T^3$ and $\rho_0=3q_2T^4$. For $A^{\mu}_0$ we have only one 
equation, namely charge conservation eq. (\ref{a6}). In terms of hydrodynamic variables 
it reads
\begin{equation}
q'_0+q_0 u^{\mu}_{,\mu}+e\tau\left(L_{0,\mu}\zeta^{\mu}+L_0\zeta^{\mu}_{,\mu}\right)
=0\label{b33}
\end{equation}
where $'\equiv u^{\mu}\partial_{\mu}$. The equations for $A^{\mu\nu}_0$ are the EMT 
conservation eqs. (\ref{a7}). Projected along $u^{\nu}$
it gives
\begin{equation}
-\rho'_0-\frac 43\rho_0u^{\mu}_{,\mu}-\frac12\tau\eta_0T^5\zeta^{\mu\nu}\sigma_{\mu\nu}
=e^2\tau L_0F_{\nu\rho}u^{\nu}\zeta^{\rho}
\label{b35}
\end{equation}
where
\begin{equation}
\sigma^{\mu\nu}=h^{\mu\rho}h^{\nu\lambda}\left[u_{\rho ,\lambda}+u_{\lambda,\rho}-
\frac 23 h_{\rho\lambda}u^{\sigma}_{,\sigma}\right]
\label{b36}
\end{equation}
is the shear tensor, and the projection orthogonal to $u^{\nu}$ yields
\begin{equation}
\frac 43 \rho_0\left(u^{\nu}\right)'+\frac13 h^{\mu\nu}\rho_{0,\mu}
+\tau\left(\eta_{0}T^5\right)_{,\mu}\zeta^{\mu\nu}+\tau\eta_0T^5 h^{\nu}_{\rho}
\zeta^{\mu\rho}_{,\mu}
=e h^{\mu\nu}F_{\mu\rho}\left(q_0 u^{\rho}+e\tau L_0\zeta^{\rho}\right)\label{b37}
\end{equation}
$A_1^{\mu\nu}$ and $A_{1}^{\mu\nu\rho}$ provide the necessary supplementary equations.
For $A_1^{\mu\nu}$ we only need its spatial projection which, after a bit of algebra 
yields
\begin{eqnarray}
0&=&\left(q_1T^4\right)_{,\mu}h^{\mu\nu}+4q_1T^4\left(u^{\nu}\right)^{\prime}
+e\tau\left(\kappa_1T^4\right)^{\prime}\zeta^{\nu}+e\kappa_1T^4\zeta^{\nu}
\nonumber\\
 &+&e\tau\kappa_1T^4 h^{\nu}_{\sigma}\left(\zeta^{\sigma}\right)^{\prime}
 +e\tau\kappa_1T^4\zeta^{\nu}u^{\mu}_{,\mu}
+\tau\left(\eta_1T^5\right)_{,\mu}\zeta^{\mu\nu}+\tau\eta_1T^5 h^{\nu}_{\sigma}
\zeta^{\mu\sigma}_{,\mu}\nonumber\\
&+&e\tau\kappa_1T^4u^{\nu}_{,\mu}\zeta^{\mu}-eq_2T^3 F^{\nu}_{\;\mu}u^{\mu}
-e^2\tau\Lambda_1T^3 h^{\nu}_{\rho}F^{\rho}_{\;\mu}\zeta^{\mu}\label{b41}
\end{eqnarray}
If we only keep the last term in the first line and the term involving the electric 
field in the last line we see this is a generalized Ohm's law.

Finally, the traceless, doubly transverse projection of the equation for $A_{1}^{\mu\nu\rho}$ 
reads
\begin{eqnarray}
-4q_2T^6\zeta^{\alpha\beta}&=&  4q_2T^5h^{\rho\beta}h^{\alpha\nu}\left[u_{\nu ,\rho} +
u_{\rho,\nu} - \frac{2}{3} h_{\rho\nu}u^{\mu}_{,\mu}\right] \nonumber\\
&-&\frac{e\tau}{10}\eta_{1,\mu}\left[h^{\mu\alpha}h^{\beta}_{\rho}\zeta^{\rho} +
 h^{\mu\beta}h^{\alpha}_{\nu}\zeta^{\nu}
-\frac{2}{3}h^{\alpha\beta}\zeta^{\mu}\right]\nonumber\\
&-& \frac{e\tau}{2}q_1 T^5h^{\alpha}_{\nu}h^{\beta}_{\rho}\left[\frac{6}{5}
u^{\nu\prime}\zeta^{\rho}
+\frac{6}{5}u^{\rho\prime}\zeta^{\nu}-\frac{4}{5}h^{\nu\rho}u^{\mu\prime}\zeta_{\mu}
\right.\nonumber\\
&+&\left.\frac{1}{5}\left(h^{\mu\nu}\zeta^{\rho}_{,\mu}+h^{\mu\rho}\zeta^{\nu}_{,\mu}
- \frac{2}{3} h^{\nu\rho}\zeta^{\mu}_{,\mu}\right)\right]\nonumber\\
&+& 4q_2\tau T^6h^{\alpha}_{\nu}h^{\beta}_{\rho}\left[\zeta^{\mu\nu}u^{\rho}_{,\mu} +
\zeta^{\rho\mu}u^{\nu}_{,\mu}-\frac{2}{3}h^{\nu\rho}\zeta^{\mu\sigma}u_{\sigma ,\mu}
+ \zeta^{\nu\rho}u^{\mu}_{,\mu}
+\zeta^{\nu\rho\prime}\right]\nonumber\\
&-&e^2\tau\kappa_1T^4 h^{\alpha}_{\nu}h^{\beta}_{\rho}u^{\mu}\left[ F^{\nu}_{~\mu}\zeta^{\rho} 
+
F^{\rho}_{~\mu}\zeta^{\nu}-\frac{2}{3}h^{\nu\rho}F_{\sigma\mu}\zeta^{\sigma}\right]
\nonumber\\
&-&\frac{4}{5}e q_1T^5\tau h^{\alpha}_{\nu}h^{\beta}_{\rho}\left[F^{\nu}_{~\mu}
\zeta^{\mu\rho}
+F^{\rho}_{~\mu}\zeta^{\mu\nu}\right]\label{b43b}
\end{eqnarray}
This may be converted into a Maxwell-Cattaneo equation\cite{JoPr89} for $\Pi^{\mu\nu}$;
 we do not need to go into this conversion in detail, as we shall adopt $\zeta^{\mu\nu}$ 
as a degree of freedom on its own.

Equations (\ref{b33}), (\ref{b35}), (\ref{b37}), (\ref{b41}) and (\ref{b43b}) together 
with the Maxwell equations (\ref{meq}) constitute our magnetohydrodynamic model. They 
break down for large anisotropies, as they do not ensure that the pressures remain 
positive, but within its range of validity they fully capture nonlinearities coming 
from the convective derivative terms and from direct coupling of the hydrodynamic variables 
to the electromagnetic fields in the equations of motion. These are the only nonlinearities 
in the usual magnetohydrodynamics, where dissipative terms are assumed to be linear.

\section{Perturbation theory}\label{pt}
To test the power of the formalism, we focus on the specially important case of 
transverse perturbations of a homogeneous anisotropic background.
The motivation behind this choice is that, since Weibel's seminal paper \cite{weibel59} 
transverse instabilities were widely studied and consequently
we can easily compare our results with the ones from different approaches to the problem. 
Another reason we can mention is that those instabilities
are found in the RHIC's experiments, where the background configuration is extremely oblate. 
As we are considering an Abelian plasma, we shall not 
be rigorously describing RHIC's instabilities, but show that it is possible to consistently 
study them without having to start from kinetic theory.

To implement our perturbative scheme, we write $u^{\mu}=u^{\mu}_0+v^{\mu}$ and 
$\zeta^{\mu\nu}=\zeta_0^{\mu\nu}+z^{\mu\nu}$; besides we consider 
$F^{\mu\nu}$ and $\zeta^{\mu}$ to be zero in the background, i.e., the electromagnetic 
variables are pure perturbations. To study the emergence of 
transverse instabilities we assume that the space-time dependence of all quantities is 
of the form $e^{st+ikz}$ and that $\zeta_0^{\mu\nu}=
\mathrm{diag}\left(\zeta_0,\zeta_0,-2\zeta_0\right)$, i.e., we consider that the pressure 
is the same along $x$ and $y$ but different along $z$. 

For an anisotropic but axisymmetric state, anisotropy is described by a dimensionless 
parameter $\tau T\zeta_0$, where $\tau$ is a characteristic 
relaxation time, $T$ is the temperature and $\zeta_0$ is an eigenvalue of the tensor 
$\zeta^{\mu\nu}$ introduced in eq. (\ref{a22}) above. 
We see from expr. (\ref{b9}) that our formalism breaks down unless $-5/4 \le \tau 
T\zeta_0\le 5/8$, as it predicts negative pressures when those 
limits are breached. However, preliminary calculations show that it remains reliable 
almost up to those boundaries \cite{MAEC17}. For this reason,
we believe these hydrodynamic equations are a valid generalization of MHD to the 
relativistic regime, and that any consistent relativistic 
dynamics of real fluids will converge to this formalism within its range of validity.

For the transverse waves, the only nonzero quantities are $v^a$, $\zeta^a$, $z^{a3}$, 
$F^{a0}$ and $F^{a3}$, with $a=x,y$. Observing that 
$\left[\tau\right]=T^{-1}$, $\left[v^a\right]=\left[z^{a3}\right]=T^0$, $\left[s\right]=
\left[k\right]=\left[\zeta^a\right]=T$ and 
$\left[F^{a0}\right]=\left[F^{a0}\right]=T^2$, there is no loss of generality in setting 
$T=1$. We also write all the coefficients in terms  
$Q = q_1/q_2$. Replacing the above defined quantities into eqs. (\ref{b33}), (\ref{b35}),
(\ref{b37}), (\ref{b41}) and (\ref{b43b})
and supplementing the system with the Maxwell equations (\ref{meq}), we obtain to first 
order in the perturbations:
\begin{eqnarray}
sv^{a}+\frac{1}{5}ik\tau z^{a3}+\frac14eQ  F^{a0} &=&0\label{c13}\\
4Qsv^{a}+e\frac1{2}\left(1-Q^2\right)\left(\tau s+1\right)\zeta^{a}+\frac 45ikQ\tau 
z^{a3}+e F^{a0}&=&0\label{c14}\\
4ik\left(1-2\tau \zeta_0\right)v^a -\frac{e\tau}{10}Q  ik \zeta^a + 4 \left(1+s\tau\right)
z^{a3}+
\frac{12}{5}e\tau Q  F^{a3}\zeta_0&=&0
\label{c15}\\
-4\pi eQv^{a}-\frac16\pi e^2\tau \left[4-3Q^2\right] \zeta^{a}+\frac1{q_2}sF^{a0}+
\frac1{q_2}ikF^{a3}&=&0\label{c16}\\
ikF^{a0}+sF^{a3}&=&0\label{c17}
\end{eqnarray}
Using the first equation we transform the second into a covariant Ohm's law
\begin{equation}
e\left(1-Q^2\right)\left[\frac1{2}\left(\tau s+1\right)\zeta^{a}+ F^{a0}\right]=0 
\label{c18}
\end{equation}
We use Faraday's law (\ref{c17}) and expr. (\ref{c18}) to write the electric field 
$F^{a0}$ and $\zeta^a$ in terms of the magnetic 
field $F^{a3}$. The above system then reduces to
\begin{eqnarray}
sv^{a}+\frac{1}{5}ik\tau z^{a3}+\frac{ieQs}{4k}  F^{a3} &=&0\label{c13b}\\
4ik\left(1-2\tau \zeta_0\right)v^a + 4 \left(1+s\tau\right)z^{a3} +\frac e5 
Q\left\{\frac{\tau s}{\left(1+\tau s\right)} +
{12}\tau \zeta_0 \right\} F^{a3}&=&0
\label{c15b}\\
4\pi i eQv^{a}+\frac{1}{q_2k}\left\{\frac13\pi e^2q_2 \left[4-3Q^2\right] 
\frac{\tau s}{1+\tau s}+\left[s^2+k^2\right]\right\}F^{a3}&=&0\label{c16b}
\end{eqnarray}
The normal modes are obtained in the usual way, by setting the determinant of the 
coefficients of system (\ref{c13b})-(\ref{c16b})
equal to zero. Considering the variables in the order $\left(v^a, z^{a3}, F^{a3} \right)$, multiplying the resulting determinant by
$q_2$ and by $\tau^3$, calling $\pi e^2\tau^2 q_2=\varpi$, $\tau s=\sigma$ and 
$\tau k=\kappa$, the dispersion relation reads
\begin{eqnarray}
0&=& \sigma^5 + 2\sigma^4 + \left\{\frac{4\varpi}{3} + 1 
+ \frac{\kappa^2}{5}\left(6-2\tau\zeta_0\right)\right\} \sigma^3 
+ \left\{\frac{4\varpi}{3} +
\frac{\kappa^2}{5}\left(11-2\tau\zeta_0\right)+ \varpi Q^2\right\}\sigma^2 
\nonumber\\
&+& \left\{\frac{\varpi\kappa^2}{15}
\left(4-3Q^2\right)\left(1-2\tau\zeta_0\right)
+ \kappa^2 + \frac{\kappa^4}{5}\left(1-2\tau\zeta_0\right)-\frac{\varpi\kappa^2Q^2}{25}
\left(1 + 12\tau\zeta_0\right) + \varpi Q^2\right\} \sigma\nonumber\\
&+& \frac{\kappa^4}{5}\left(1-2\tau\zeta_0\right) - \frac{12}{25} \varpi\kappa^2 
Q^2\tau\zeta_0\label{nm-1}
\end{eqnarray}
To avoid the possibility of an unphysical background with negative pressures we assume 
$\tau\zeta_0\ll 1$, whereby this relation 
simplifies to
\begin{eqnarray}
0&=& \sigma^5 + 2\sigma^4 + \left\{ 1 + \frac{4\varpi}{3} + \frac{6\kappa^2}{5}\right\} 
\sigma^3 
+ \left\{\frac{4\varpi}{3} + \frac{11\kappa^2}{5}+ \varpi Q^2\right\}\sigma^2 \nonumber\\
&+& \left\{\frac{4\varpi \kappa^2}{15}  + \kappa^2 + \frac{\kappa^4}{5}-\frac{6\varpi 
\kappa^2Q^2}{25}
+ \varpi Q^2\right\} \sigma\nonumber\\
&+& \frac{\kappa^4}{5} - \frac{12}{25} \varpi\kappa^2 Q^2\tau\zeta_0\label{nm-2}
\end{eqnarray}
Since $Q^2\leq 1$ the linear term is always positive. Therefore the necessary and 
sufficient condition for the emergence
of instabilities is the independent term to be negative. This gives the condition for 
unstable modes
\begin{equation}
 \kappa^2 \leq \kappa_{max}^2 = \frac{12}{5}\varpi Q^2\tau\zeta_0\label{nm-3}
\end{equation}
which is only within the rage of validity of our model provided that
\begin{equation}
 \frac{\kappa^2}{\varpi Q^2}\ll 1 \label{nm-4}
\end{equation}
We must now find the interval of possible values of $\sigma$ consistent with the bounds 
found above. To this purpose we
first discuss the dependence of the solution with respect to $\kappa^2$. If $\sigma_{\kappa}$ 
is the value of the root for 
a given value of $\kappa^2$ we rewrite expr. (\ref{nm-2}) as a polynomial in $\kappa$ 
as
\begin{equation}
 a\left[\sigma_{\kappa}\right] \kappa^4 + b\left[\sigma_{\kappa}\right] \kappa^2 
+ c\left[\sigma_{\kappa}\right] =0 \label{nm-5}
\end{equation}
where
\begin{eqnarray}
 a\left[\sigma_{\kappa}\right] &=& \frac{1}{5}\left( 1 +  \sigma_{\kappa} \right)\nonumber\\
 b\left[\sigma_{\kappa}\right] &=& \frac{6}{5}\sigma_{\kappa}^3 + \frac{11}{5}\sigma_{\kappa}^2 
+
 \left[ 1 + \frac{4}{15}\varpi - \frac{6}{25}\varpi Q^2\right]\sigma_{\kappa} - 
\frac{1}{5}\kappa_{max}^2 \label{nm-6}\\
 c\left[\sigma_{\kappa}\right] &=&\sigma_{\kappa}\left[ \sigma_{\kappa}^4 + 
2 \sigma_{\kappa}^3 + \left[ 1 + \frac{4}{3}\varpi\right]\sigma_{\kappa}^2
 +\left[\frac{4}{3} + Q^2\right]\varpi \sigma_{\kappa} + \varpi Q^2 \right]\nonumber
\end{eqnarray}
It is easily seen that $\sigma_{\kappa} = 0$ corresponds to either $\kappa = 0$ or else 
$\kappa = \kappa_{max}$. Observe that
$a\left[\sigma_{\kappa}\right]$ and $c\left[\sigma_{\kappa}\right]$ are always positive 
definite, while $b\left[\sigma_{\kappa}\right]$
is negative at $\sigma_{\kappa} = 0$ and then grows, eventually reaching 0. Therefore at 
$\sigma_{\kappa} = 0$ we have that
$b\left[\sigma_{\kappa}\right]^2 > 4a\left[\sigma_{\kappa}\right]c\left[\sigma_{\kappa}\right]$
 but there exists a critical value
$\sigma_{\kappa c}$ such that $b\left[\sigma_{\kappa}\right]^2 = 
4a\left[\sigma_{\kappa}\right]c\left[\sigma_{\kappa}\right]$ and 
for which there is only one possible value of $\kappa^2$, namely
\begin{equation}
 \kappa_c^2 = \sqrt{\frac{c\left[\sigma_{\kappa c}\right]}{a\left[\sigma_{\kappa c}\right]}}
\label{nm-7}
\end{equation}
Clearly, $\sigma_{\kappa c}$ must be smaller than the root of $b\left[\sigma_{\kappa }\right]$,
 which in turn is smaller than
$\kappa_{max}^2/5$. In this way we obtained an upper bound for the possible values of $\sigma$,
 namely 
\begin{equation}
\sigma \leq  \frac{1}{5}\kappa_{max}^2\label{nm-8}
\end{equation}

We must now perform a similar analysis with respect to the parameter $\varpi$. For a given
 $\kappa$ we have
\begin{equation}
 S\left[\sigma\right]\varpi + R\left[\sigma\right]=0\label{nm-9}
\end{equation}
with
\begin{eqnarray}
 S\left[\sigma\right] &=& \frac{4}{3}\sigma^3 + \left[\frac{4}{3}+Q^2\right]\sigma^2 +
 \left[\frac{4}{15}\kappa^2 -
 \frac{6}{25}Q^2\kappa^2 + Q^2\right]\sigma - \frac{12}{25}\kappa^2Q^2\tau\zeta_0
\label{nm-10a}\\
 R\left[\sigma\right] &=& \sigma^5 + 2\sigma^4 + \left[1+\frac{6}{5}\kappa^2\right]\sigma^3
 +\left[\kappa^2 + \frac{\kappa^4}{5}\right]\sigma
 + \frac{\kappa^4}{5}\label{nm-10b}
\end{eqnarray}
As $R\left[\sigma\right]$ is clearly positive, $S\left[\sigma\right]$ must be negative for 
eq. (\ref{nm-9}) be zero. 
For a given $\kappa$, the instability exists only if $\varpi$ exceeds the value for which 
$\kappa^2 = \kappa_{max}^2$. When 
$\varpi \rightarrow \infty$, $\sigma$ approaches the lowest positive root of 
$S\left[\sigma\right]$.

As the derivatives of $R\left[\sigma\right]$ and $S\left[\sigma\right]$ are both positive, 
the roots of the polynomial 
(\ref{nm-9}) are growing functions of $\varpi$. For this reason, the asymptotic limit of 
$\varpi\to\infty$ provides an upper
bound for the roots at finite values of $\varpi$.
We thus obtain a more strict bound on $\sigma$, namely
\begin{equation}
\sigma\le\frac{\frac{12}{25} \kappa^2 Q^2\tau\zeta_0}{\frac{4\sigma^2}{3} +\left[\frac{4}{3}
+ Q^2\right]\sigma +4\frac{\kappa^2}{15}-
6\frac{\kappa^2Q^2}{25}+ Q^2}\le 18Q^2\tau\zeta_0\label{nm-11}
\end{equation}
Since within the range of validity of our model this implies that $\sigma\ll 1$, it is enough 
to keep only the independent and linear terms in
expr. (\ref{nm-2}). Therefore we obtain the following dispersion relation
\begin{equation}
\sigma_{\kappa} = \frac{\left(\kappa_{max}^2 - \kappa^2\right)\kappa^2}{\frac{4}{3}\varpi
\kappa^2+5\kappa^2
+\kappa^4 - \frac{6}{5}\varpi \kappa^2 Q^2 + 5\varpi Q^2}\label{nm-12}
\end{equation}
which also gives an upper bound for the exact time constant. Observe that for $Q^2=0$ there 
are no unstable modes, i.e., all values of
$s$ are negative. For $Q^2\not=0$, instabilities, namely $s>0$ will exist only for 
$0<\kappa^2<\kappa_{max}^2$, since the denominator
of. eq. (\ref{nm-12}) is positive.
In the following figures we plot the dispersion relation for different values of the parameters 
$Q$, $\tau$ and $\zeta_0$, consistent with
the above quoted intervals of validty and with bound (\ref{nm-11}). In the three cases, 
the higher the values of the parameters, the larger the 
$\kappa$ interval for instabilities, 
as expected. These features are in agreement with previous results found in the literature 
\cite{YooDav87,SSGT06,Hon04,Yoon07}.

\begin{figure}[!htb]
 \includegraphics[scale=0.7]{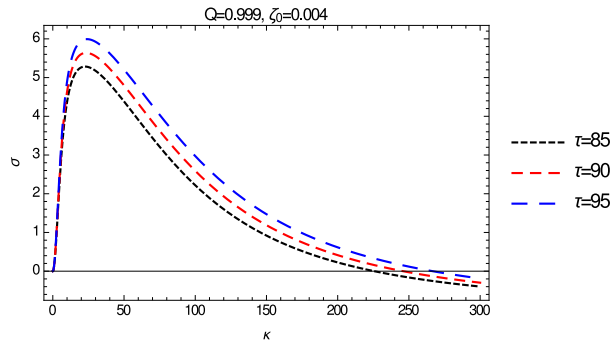} 
 \caption{\label{sQ} Plot of $\sigma_{\kappa}$ as a function of $\kappa$ from expr. 
(\ref{nm-12}), for fixed values $\zeta_0=0.004$, 
 $Q= 0.999$, and $\tau = 85$ ($\varpi = 62,459.5 $) (black, short-dashed), $\tau = 90$ 
($\varpi = 70,023.8 $) (red, medium-dashed) 
 and $\tau = 95$ ($\varpi = 78,020.3 $) (blue,long- dashed).  Larger values of $\tau$ allow for 
more unstable modes, as expected.}
 \label{fig-1}
\end{figure}

\begin{figure}[!htb]
 \includegraphics[scale=0.7]{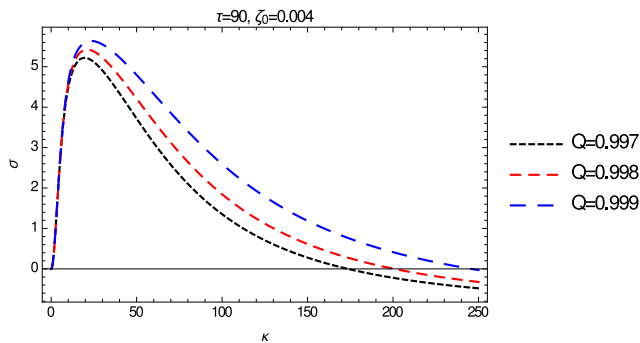} 
 \caption{\label{stau} Plot of $\sigma_{\kappa}$ as a function of $\kappa$ from expr. 
(\ref{nm-12}), for fixed values $\tau = 90$, $\zeta_0=0.004$, 
 $Q= 0.997$ ($\varpi = 35,011.9 $) (black, short-dashed), $Q= 0.998$ ($\varpi = 46,682.5 $) 
(red, medium-dashed) and $Q= 0.999$ 
 ($\varpi = 70,023.8 $) (blue, long-dashed). As expected, a larger excess of particles 
over antiparticles allows for more unstable modes.}
 \label{fig-2}
\end{figure}

\begin{figure}[!htb]
 \includegraphics[scale=0.7]{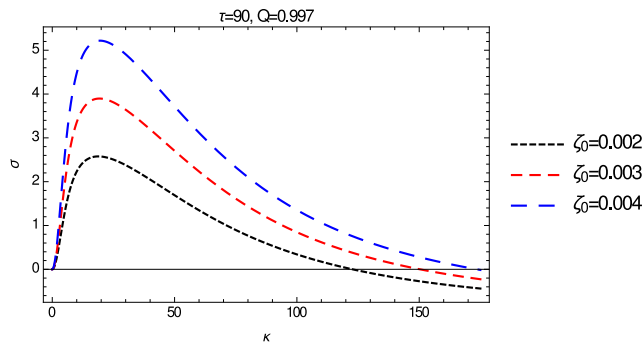} 
 \caption{\label{szeta} Plot of $\sigma_{\kappa}$ as a function of $\kappa$ from expr. 
(\ref{nm-12}), for fixed values $\tau=90$, 
 $Q = 0,997$ ($\varpi = 35,011.9 $) and $\zeta_0=0.002$ (black, short- dashed), 
$\zeta_0=0.003$ (red, medium-dashed) and $\zeta_0=0.004$ (blue, long-dashed). 
 Again, larger values of the background anisotropy allow for more unstable modes.}
 \label{fig-3}
\end{figure}

One last consideration concerns the fact that the magnetic field grows when the system 
is in the only unstable mode. 
This can be seen from eq. (\ref{c16b}), because if $F^{a3}$ would be zero, then so is $v^a$, 
and all amplitudes would vanish.

\section{Conclusions}\label{cl}

In this manuscript we have built a minimal magnetohydrodynamic formalism to describe 
highly relativistic dissipative plasmas and their instabilities
in a unified way. For consistency at microscopic and macroscopic levels, we anchored 
the hydrodynamics to kinetic theory by writing all tensors 
of the model as moments of a 1pdf, and their corresponding evolution equations as moments 
of a Vlasov-Boltzmann equation. For the collision
integral, we used the Anderson-Witting prescription, which is a linear function of the 
non-equilibrium part of the 1pdf. 
This choice proved to describe more accurately highly relativistic systems than the BGK 
ansatz, as explained in Section \ref{int}. We built the 
non-equilibrium 1pdf by introducing two new tensors, $\zeta^{\mu}$ and $\zeta^{\mu\nu}$ 
in such a way that the conduction currents and 
viscous stresses are linear on them.
This simplifies the mathematics at the price of enforcing positivity of the pressures; 
however, the resulting formalism contains all the 
nonlinearities already present in the usual MHD, namely those coming from convective 
terms and from the coupling to the electromagnetic fields.

We applied our formalism to analyze transverse normal modes around an anisotropic 
background. We found a dispersion relation, eq. (\ref{nm-12})
consistent with the approximations made. This relation describes instabilities in the 
long wavelength range, with features that are in agreement with 
those found in previous works \cite{YooDav87,SSGT06,Hon04,Yoon07}. Our model is robust 
in the sense that no fine-tunning was needed to get these results.

In other words, we provided a check that pure hydrodynamic schemes are rich enough to 
describe the essential features of a anisotropic 
instabilities. We observe that there are in the literature mixed analyses in which the 
linearized fluctuations around an anisotropic solution 
to kinetic theory are described in hydrodynamic ways\cite{CJPR13}.

To the best of our knowledge, this is the first time a set of hydrodynamic equations 
is presented that describe both the background
and the fluctuations. In last analysis, the usefulness of having a purely hydrodynamic 
theory is that it should make much easier to test it
against experimental results, to implement numerical simulations and to follow the 
evolution of the instability beyond the linearized 
approximation 
\cite{ArMo06,RSA08,RS10,IRS11,ARS13b,AKLN14,FIWi11,Fuk13,Khac08,CaRe10}. We expect to 
report on this last issue in the near future as well as on 
the extension of our formalism to non-Abelian relativistic plasmas and also a full 
comparison between hydrodynamic instabilities and the more 
detailed description that follows from kinetic theory with an Anderson-Witting collision 
term 
\cite{and-witt1,and-witt2,TI10,epv-2,SSGT06}.

\section*{Acknowledgments}
 
We thank P. Romatschke, M. Strickland and T. Christen for useful comments. E. C. 
acknowledges support from ANPCyT, CONICET and the University of 
Buenos Aires. A. K. thanks financial support from FAPESB grant 
AUXPE-FAPESB-3336/2014/Processo 23038.007210/2014-19 and FAPESB grant
FAPESB-PVE-015/2015/Processo PET0013/2016, and Universidade Estadual de Santa Cruz.

\end{document}